\documentclass[12pt,preprint]{aastex}
\usepackage{hyperref}
\usepackage{graphicx,epsfig,wrapfig,paralist,sidecap} 
\usepackage{longtable}

\newcommand{\be}{\begin{equation}}
\newcommand{\ee}{\end{equation}}
\newcommand{\ba}{\begin{align}}
\newcommand{\ea}{\end{align}}

\newcommand{\gae}{\lower 2pt \hbox{$\, \buildrel {\scriptstyle >}\over {\scriptstyle
\sim}\,$}}
\newcommand{\lae}{\lower 2pt \hbox{$\, \buildrel {\scriptstyle <}\over {\scriptstyle
\sim}\,$}}
\newcommand{\aprop}{\lower 2pt \hbox{$\, \buildrel {\scriptstyle \propto}\over 
   {\scriptstyle \sim}\,$}}

\usepackage{natbib}
\begin{document}

\title{Some Implications of inverse-Compton Scattering of Hot Cocoon Radiation by relativistic jets in Gamma-Ray Bursts}

\author%[Kumar \& Smoot]
{ Pawan Kumar$^1$ and George F. Smoot$^{2,3}$
\thanks {E-mail:  pk@astro.as.utexas.edu, gfsmoot@lbl.gov} \\
$^{1}$Department of Astronomy, University of Texas at Austin, Austin, TX 78712, USA \\
$^{2}$PCCP; APC, Universit\' e Paris Diderot, Universit\'e Sorbonne Paris Cit\'e,  75013 France\\
$^{3}$BCCP; LBNL \& Physics Dept. University of California at Berkeley  CA 94720, USA}

\date{Accepted; Received; in original form 2014 February 11}

%\pubyear{2014}

%\keywords{radiation mechanisms: non-thermal - methods: analytical  
 %- gamma-rays: bursts, hot cocoon, theory}

\maketitle

%\begin{abstract}
{\it Abstract:} Long Gamma-Ray Bursts (GRB) relativistic jets are surrounded 
by hot cocoons which confine jets during their punch out from the progenitor
star. These cocoons are copious sources of X-ray photons that can be and 
are inverse-Compton (IC) scattered to MeV--GeV energies by electrons in 
the relativistic jet. We provide detailed estimates for IC flux resulting
from various interactions between X-ray photons and the relativistic jet, 
and describe what we can learn about GRB jets and progenitor stars from 
the detection (or an upper limit) of these IC scattered photons.

%r\end{abstract}

\section{Introduction}

There is evidence that long duration gamma-ray bursts (GRBs) are produced 
when a massive star dies at the end of its nuclear burning life and its 
core collapses to a neutron star or a blackhole (e.g. Galama et al. 1998, 
Hjorth et al. 2003, Stanek et al. 2003, Modjaz et al. 2006, 
Campana et al. 2006, Starling et al. 2011, Sparre et al. 2011, 
Melandri et al. 2012). 
The newly formed compact object at the center of the progenitor star produces 
a pair of relativistic jets that make their way out of the star along polar 
regions. Punching their way to the stellar surface these jets shock heat 
the material they encounter and push it both sideways and along the jet's 
direction. Therefore, the jet is surrounded by this shock heated plasma, 
or a hot cocoon, which provides collimation for it (see fig. 1). 
The total amount of thermal energy in the cocoon is equal 
to the work done by the jet on stellar material it encounters while inside 
the star and that is estimated to be of order $L_j t_*\sim 10^{51}$erg --- 
where $L_j$ is the jet luminosity and $t_*$ is the travel time for the jet 
inside the star at sub-relativistic speed (e.g. Meszaros \& Rees, 2001;
Ramirez-Ruiz et al. 2002; Matzner, 2003).
We are here making the rough assumption that the bulk of the work goes 
to thermal energy though a significant amount goes into the cocoon forward 
(in jet direction) momentum.

\begin{figure}[ht!]
\centering
\includegraphics[scale=0.17]{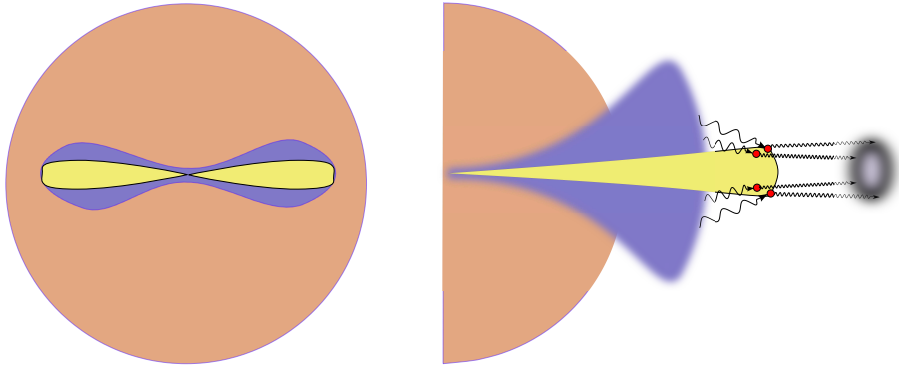}
\caption{Schematic sketch of a hot cocoon surrounding the jet while the jet is 
still inside the star is shown on the left side of this figure. The sketch 
to the right shows half of the system (the hemisphere with right to
the center jet) at a later time when the jet and the cocoon 
have punched through the surface of GRB progenitor star. Thermal radiation 
from the cocoon is inverse-Compton scattered by the same jet, and by any 
relativistic jet produced by the central engine at a later time when
the {\it late jet} emerges above the cocoon surface. The IC scattered
photons form a halo peaked near the edge of the jet and roughly as wide 
as the jet.
\label{fig1} }
\end{figure}

The jet and the cocoon emerge from the stellar surface more or less at the
same time. The central engine of long-GRBs remains active for $\sim$10s to 
10$^5$s after the jet emerges from the stellar surface (e.g. Burrows et al. 
2005, Chincarini et al. 2007, 2010). The isotropic equivalent of luminosity 
from the cocoon at its peak, shortly after it emerges from the stellar 
surface, is of order 10$^{49}$ erg s$^{-1}$ at a few keV 
(Ramirez-Ruiz et al. 2002). The cocoon luminosity decreases 
slowly with time for about 10$^3$s and then drops to zero rapidly when 
the cocoon becomes transparent to Thomson scattering of photons.
Any relativistic jet that is launched after the cocoon breaks through the 
stellar surface is expected to encounter this intense radiation
field when it reaches the cocoon photosphere, and will produce a short-lived 
bright pulse of inverse-Compton scattered photons which 
form a halo peaked near the edge of the jet and roughly as wide as 
the jet (fig. 1)\footnote{The
jet while inside the cocoon is shielded from the IC drag of its intense 
radiation by an optically thick layer of electron-positron plasma
that is created at the interface of the jet and cocoon. These $e^\pm$
pairs are produced by the collision of thermal photons from the cocoon 
with IC scattered photons at this interface (Ceccobello \& Kumar, 2014).}.
Furthermore, radiation from the cocoon is Thomson scattered by electrons
in the wind of GRB progenitor star, and some of these wind-scattered photons 
collide with the relativistic jet at large angles to produce an intense pulse
of high energy inverse-Compton radiation.

We provide in this work an estimate of IC flux from these different 
interactions of cocoon radiation with the GRB relativistic jet, and suggest
how the detection of this high energy radiation (or an upper limit) can be 
used to constrain GRB jet and progenitor star properties.

In section 2 we describe cocoon dynamics and radiation which follows 
closely the work of Ramirez-Ruiz et al. (2002) and Matzner (2003), and 
is included here for the sake of completeness and ease to follow the 
rest of the paper. Section 3 describes the IC scattering of cocoon 
radiation by a relativistic jet directly, and via an intermediate process 
of scattering first by electrons in the circum-burst medium (CBM).
Application to GRBs is provided in \S4.

\section{Cocoon dynamics and radiation}

We describe in this section the propagation of a relativistic jet through
a star and how it shocks and pushes sideways stellar material it encounters
on its way to the surface to evacuate a cavity in the polar region. This
process creates a hot cocoon of plasma that encapsulates the jet. The 
properties of the hot
cocoon and its radiation are discussed separately in two subsections.
Much of the discussion in this section closely follows the work of 
Ramirez-Ruiz et al. (2002) and Matzner (2003).

\subsection{Dynamics of cocoon}

Let us consider a jet of luminosity $L_j$, speed $v_j$, and Lorentz factor
$\Gamma_j$ produced by the central engine of a GRB. While the jet moves 
through the star it drives a shock wave through the stellar envelope and 
that in turn reacts back on the jet and slows down a section of it 
near the head via a reverse-shock traveling into the jet. 

The speed of the jet head, $v_h$, can be calculated from ram pressure 
balance in the radial direction of the unshocked jet and stellar plasma
as viewed from the rest frame of the jet-head:
\be
 \rho_j c^2 (\Gamma_j^2/4\Gamma_h^2) \approx \rho_a \Gamma_h^2 v_h^2,
\ee
where $\rho_j$ and $\rho_a$ are densities of the unshocked jet and the
stellar envelope respectively, and $\Gamma_j$ \& $\Gamma_h$ are the Lorentz
factors of the unshocked jet and the jet-head wrt the unshocked star (the 
Lorentz factor of the unshocked jet wrt jet-head is $\sim\Gamma_j/2\Gamma_h$).
Considering that the jet luminosity at the stellar surface can be written as
\be
   L_j = \pi \theta_j^2 R_*^2 \rho_j \Gamma_j^2 c^3,
  \label{L_j}
\ee
and the total mass of the swept up gas by the jet is
\be
   m_c \sim \pi \theta_j^2 \rho_a R_*^3,
  \label{m_c}
\ee
we obtain
\be
  2 \Gamma_h^2 v_h \sim \left[ {R_* L_j\over c m_c}\right]^{1/2},
  \label{vh}
\ee
where $\theta_j$ is jet opening angle.

The work done by the jet on the stellar material can be obtained from
momentum flux conservation and from the fact that the energy flux for 
relativistic outflows is $c$ times the momentum flux whereas for the 
sub-relativistic swept-up gas it is smaller by a factor $v_h/c$. The 
difference in the energy flux for the jet-head and the unshocked jet 
gives the rate at which energy is deposited in the cocoon, and
this way we obtain the total energy in the cocoon to be
\be
  E_c\sim L_j \left[ 1 - v_h/c\right] R_*/v_h \quad{\rm or}\quad 
    E_c\sim L_j R_*/v_h \quad{\rm for}\;\; v_h/c \ll 1.
  \label{E_c1}
\ee

With this expression for energy in the cocoon we can simplify equation
(\ref{vh}) further:
\be
  4 \Gamma_h^4 v_h \sim c \Gamma_c
  \label{vh1}
\ee
where
\be
   \Gamma_c \equiv {E_c\over m_c c^2}
  \label{eta_c}
\ee
is the terminal Lorentz factor of the cocoon plasma (provided that
$\Gamma_c\ge1$) after it escapes through the stellar surface and its
thermal energy is converted to bulk kinetic energy.\\
Therefore, the jet head speed is sub-relativistic when $\Gamma_c< 4$ and
is given by
\be
   v_h \sim c\Gamma_c/4.
   \label{vh2a}
\ee
For $\Gamma_c>4$, the jet head speed is mildly relativistic and its Lorentz 
factor is given by
\be
   \Gamma_h \sim (\Gamma_c/4)^{1/4}.
   \label{vh2b}
\ee
The expansion speed of the cocoon in the
direction perpendicular to its surface, $v_c$, is determined by equating
the ram pressure with the thermal pressure inside the cocoon ($p_c$),
i.e.
\be
  v_c = (p_c/\rho_a)^{1/2}.
  \label{vc1}
\ee
The average thermal pressure inside the cocoon is approximately
\be
  p_c = \rho_a v_c^2 \sim {E_c\over 3 V_{c}}.
  \label{p_c}
\ee
where
\be
 V_{c}\sim \pi t^3 v_h v_c^2/3 \sim R_*^3 (v_c/v_h)^2
 \label{cocoon-volume}
\ee
 is the volume of the cocoon. Using equations (\ref{p_c}) \& 
(\ref{cocoon-volume}) we find
\be
  v_c^4\sim {E_c v_h^2 \over 3\rho_a R_*^3} \sim \pi \theta_j^2 \Gamma_c 
     v_h^2 c^2/3.
  \label{vc2}
\ee
We made use of equations (\ref{m_c}) and (\ref{eta_c}) to derive the last 
equality.

Substituting for $v_h$ from equations (\ref{vh2a}) and (\ref{vh2b}) we obtain
\begin{equation}
 {v_c\over c} \sim \left\{
 \begin{array}{ll}
   \hskip -5pt (\pi\theta_j^2 \Gamma_c^3/48)^{1/4}  & {\rm for}\; \Gamma_c<4 \\ \\
   \hskip -5pt  (\pi\theta_j^2 \Gamma_c/3)^{1/4}     & {\rm for}\; 
    4<\Gamma_c<\theta_j^{-2}
 \end{array}
\right.
   \label{vc3}
\end{equation}
The thermal pressure of the cocoon can be obtained using equations
(\ref{vc1}) and (\ref{vc3}) and is given by
\be
  p_c \sim {L_j \over \theta_j R_*^2 c (3\pi\Gamma_c)^{1/2} },
  \label{pc}
\ee
and its temperature is
\be
   k_B T_c = k_B (3 p_c/\sigma_a)^{1/4} \sim (7 {\rm keV})
   { \left[ L_{j,52}^{(iso)} \theta_{j,-1}\right]^{1/4} \over
  R_{*,11}^{1/2} \Gamma_{c,1}^{1/8} },
   \label{Tc}
\ee
where $\sigma_a$ is the radiation constant, and $L_j^{(iso)} = 4 
L_j/\theta_j^2$ is isotropic equivalent of GRB jet average luminosity. 
We note that the cocoon temperature has a weak dependence on jet luminosity 
and its unknown angular size while inside the star.

\begin{figure}[h!]
\centering
\includegraphics[scale=0.6]{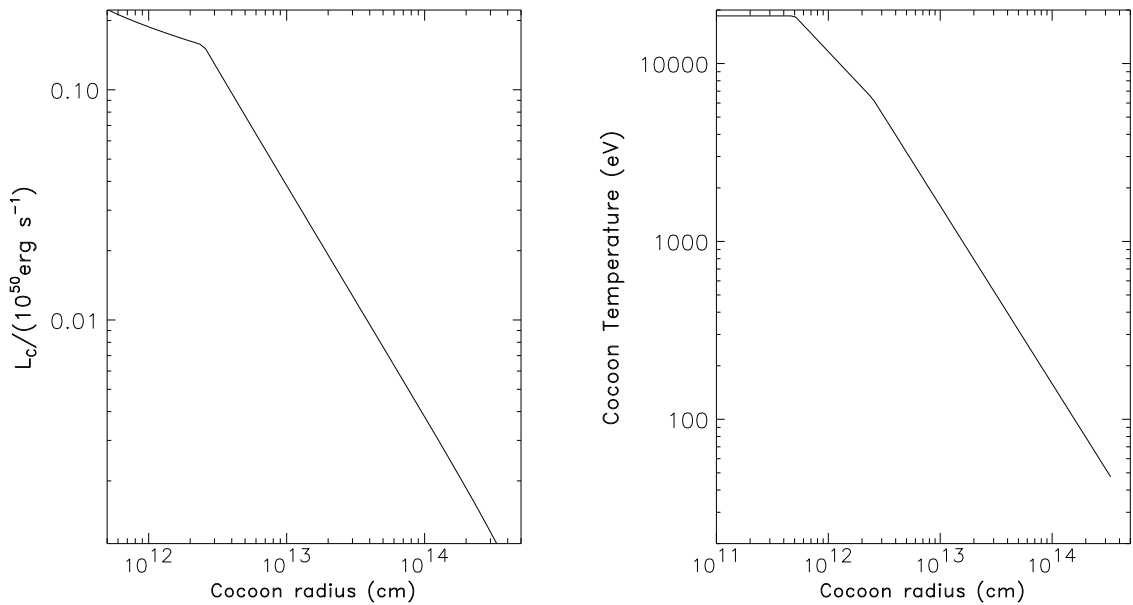}
\caption{The left panel shows the isotropic equivalent of luminosity of 
cocoon in GRB host galaxy rest frame (divided by 10$^{50}$erg/s) as
a function of its distance from the center of explosion.
The thermal energy in the cocoon is taken to be 10$^{52}$ erg 
(isotropic equivalent), its terminal Lorentz factor ($\Gamma_c$) is 5, and the 
radius of GRB progenitor star is taken to be 10$^{11}$cm. The cocoon
radiation lasts for about 10$^2$s in observer frame, and is likely 
hidden under a much brighter GRB prompt emission and its X-ray tail.
 The right panel shows the cocoon temperature in eV as a function of its
distance from the center.
   \label{cocoon-prop} }
\end{figure}

Once the cocoon punches through the stellar surface its Lorentz factor
increases linearly with radius, as is the case for any radiation dominated 
relativistic plasma, until it attains the terminal value of $\Gamma_c$.
The temperature in the observer frame during this phase of acceleration
is constant. The temperature decreases as $r^{-2/3}$ when the cocoon starts 
to coast at a constant speed of $v_c$ at $R_s \sim R_*\Gamma_c$ but before 
its radial width starts to increase linearly with distance at 
$r\gae R_*\Gamma_c^2$; the temperature declines as $r^{-1}$ for 
$r\gae R_*\Gamma_c^2$.

\subsection{Thermal radiation from cocoon}

The average number density of electrons associated with baryons in the cocoon, 
in its rest frame, is
\be
  n'_{p}(R_*) \sim {m_c\over m_p V_{c}} \sim {3 p_c \over m_p c^2 \Gamma_c}
   \sim (2\times10^{19}{\rm cm}^{-3}) { L_{j,52}^{(iso)} \theta_{j,-1} \over
  R_{*,11}^2 \Gamma_{c,1}^{3/2} }.
\ee
This density is much larger than the number density of thermal $e^\pm$ pairs 
at temperature $T_c$, 
\be
   n'_\pm = {2 (2\pi k_B m_e T_c)^{3/2} \over h^3} \exp\left(-m_e c^2/k_B T_c
   \right),
\ee
as long as $k_B T < 30$ keV.\\

The photon mean free path length in cocoon comoving frame is
\begin{equation}
    \lambda'(r) = {1\over \sigma_T n'_p(r)} \approx
      8\times10^4 {\rm cm}\, {R_{*,11}^2 \Gamma_{c,1}^{3/2} \over
     L^{(iso)}_{j,52} \theta_{j,-1} } \left[ {r\over R_*} \right]^3
    \min\left\{1, \, \max\left[R_s/r, \, \Gamma_c^{-1}\right] \right\},
\end{equation}
where $R_s = \Gamma_c R_*$ is the radius where Lorentz factor of the cocoon 
stops increasing, $R_*$ is the GRB progenitor star radius, and $\theta_j$ 
is its average opening angle during the time when it was making its way 
out of the star.

The cocoon thermal luminosity is governed by photon diffusion across 
the cocoon and is given by
\begin{equation}
    L_c^{iso}(r) = 4\pi R_*^2 \sigma_{sb} T_c^4(R_*) (r/R_s)^{2/3}
       (\lambda'(r)/c t')^{1/2}
   \label{L_cocoon}
\end{equation}
where $\sigma_{sb}$ is the Stefan-Boltzmann constant, $T_c(R_*)$ is cocoon 
temperature when it emerges from the stellar surface, and $t'=r/(c\Gamma_c)$ 
is dynamical time in cocoon rest frame.
This equation for cocoon luminosity is valid as long as the optical depth
is much larger than unity, i.e. for $\lambda'(r) \ll r/\Gamma_c$. The
luminosity and temperature are shown in fig. \ref{cocoon-prop}. 
The radius where the cocoon becomes transparent to Thomson scatterings,
i.e. $\lambda'(r) \approx r/\Gamma_c$, is given by:
\begin{equation}
  R_{tr} = \left[ {\sigma_T E_c\over 4\pi m_p c^2 \Gamma_c} \right]^{1/2},
  \label{r_ph_cocoon}
\end{equation}
where $E_c$ is the total energy of the cocoon. The optical depth of the
cocoon for $r<R_{tr}$ scales as $r^{-2}$.

The cocoon luminosity drops quickly for $r> R_{tr}$ since 
the total rate of emission due to the bremsstrahlung process 
at radius $R_{tr}$ can be shown to be rather small:
\begin{equation}
  L_{ff}(r=R_{tr}) \approx (1.3\times10^{40} {\rm erg s^{-1}}) [T_{c}(R_{tr})
/10^7{\rm K}]^{1/2} R_{tr,14} \Gamma_c^{2.5}.
\end{equation}

\section{IC scattering of cocoon radiation by jet}

\begin{figure}[ht!]
\centering
\includegraphics[scale=0.17]{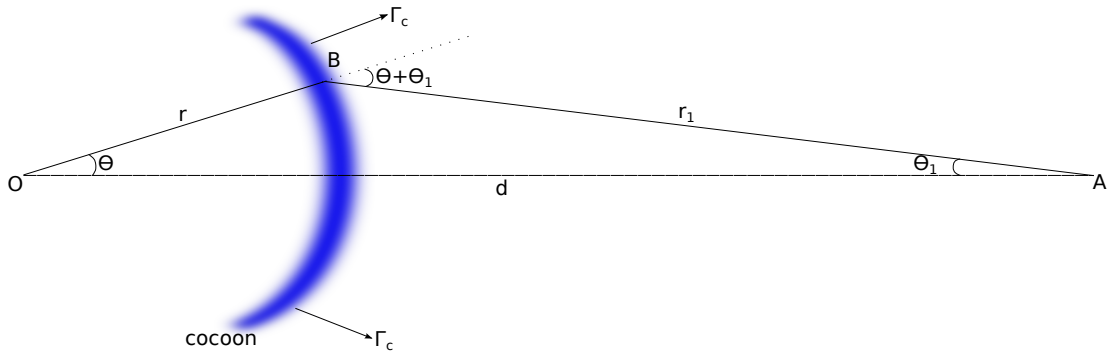}
\caption{The basic geometry of inverse-Compton scattering of thermal 
radiation from cocoon by electrons in a relativistic jet. A photon 
leaves the cocoon at $(r, \theta, t_1)$ --- point marked ``B'' in the sketch, 
and arrives at ``A'' where the relativistic jet is at time $t$. The angle 
between the photon momentum vector and jet axis is $\theta_1$. After
IC scattering the photon travels within an angle $\Gamma_j^{-1}$ of 
the radial direction because of relativistic beaming; $\Gamma_j$ is the
Lorentz factor of the jet. 
   \label{fig2} }
\end{figure}

For simplicity we will consider the cocoon, after it emerges from the
surface of the star and following a brief period of acceleration, to be a 
shell of plasma that moves away from the star at a constant speed $v_c$ 
and Lorentz factor $\Gamma_c$.
After emerging from the star the cocoon undergoes adiabatic expansion 
and converts a part of its thermal energy to kinetic energy of bulk 
motion. During this phase its Lorentz factor increases linearly with
radius and reaches the terminal value $\Gamma_c$ at $r \sim \Gamma_c R_*$. 
The distance of the cocoon from the center of the star is approximately
given by
\begin{equation}
   R_c(t) \approx R_* + t v_c.
\end{equation}
The time $t$ is measured in the GRB host galaxy rest frame (as are all
times unless specified otherwise), and its zero-point is taken to coincide 
with the emergence of the cocoon from stellar surface.

Let us assume that a relativistic jet is launched from GRB central engine 
at time $t_j$ after the cocoon punches through the surface of the star.
 The jet carries a luminosity $L_j^{(iso)}$ (isotropic equivalent),
has Lorentz factor $\Gamma_j$ and speed $v_j$.  

The time when the jet emerges above cocoon surface is given by
\begin{equation}
  t_{emerge} = {R_* + t_j v_c\over v_j - v_c},
\end{equation}
when it is at a distance $r_{emerge} = R_* + v_c t_{emerge}$ from the center.

We will not consider any inverse-Compton scattering of cocoon 
radiation by the jet until the jet punches through the cocoon photosphere 
or gets to a radius where the cocoon is transparent to photons (which
ever comes first). This is to avoid the uncertainty regarding the escape 
of photons below the photosphere through the polar cavity which is likely
to be partially filled by plasma flowing into it in between episodes
of central engine activity.

Thermal photons from the cocoon can either travel directly to the
relativistic jet and undergo inverse-Compton scattering there.
Or photons from the cocoon could first be Thomson scattering by 
electrons in the circum-burst medium (CBM) toward the jet,  and then 
undergo inverse-Compton scattering by electrons in the relativistic jet. 
We consider these two different possibilities separately in subsections 
below. It might seem that the second process is less efficient than 
the first, and
hence might not have any impact on observations. However, because of 
relativistic beaming of thermal photons from the cocoon in the forward 
radial direction, the angle between the jet axis and photons is of order 
$\Gamma_c^{-1}$ or less, and hence these IC scatterings boost photon energy 
by a factor $\sim (\Gamma_j/\Gamma_c)^2$ or less when electrons are cold 
in the jet comoving frame. 
On the other hand, thermal photons scattered first by electrons in the 
CBM can collide with the jet at larger angles and produce much higher 
energy photons. 

\subsection{Direct scattering of cocoon photons by the relativistic jet}

 The basic process considered in this section is 
sketched in figure 1. Photons from the cocoon travel directly to the
relativistic jet and there they are IC scattered by electrons in the jet. 

We consider all those photons from the cocoon that arrive at the jet at 
time $t$ when it is a distance $d$ from the center (fig. \ref{fig2} -- the point 
marked ``A''), and determine the IC luminosity when some of these photons 
are scattered by the jet. Photons leave the cocoon photosphere at angle
$\theta$, when it is at radius $r$, and meet the jet at radius $d$ 
(see fig. \ref{fig2}) at time $t$ when the following condition is satisfied:
\begin{equation}
   t = (r - R_*)/v_c + r_1/c,
\end{equation}
where 
\begin{equation}
 r_1^2 = d^2 + r^2 - 2 r d\, \cos\theta.
\end{equation}
This pair of equations determine the locus of all points in the
$(r, \theta$) plane --- the equal arrival time curve --- from which
photons arrive at ``A'' from the cocoon. We note that only a 
fraction of photons originating at the {\it equal arrival time
curve} make it to the jet. Those that travel at an angle
$(\theta +\theta_1) > \Gamma_c^{-1}$ are swept back up by the
cocoon and scattered in a different direction depending on
cocoon's optical thickness at the time of this encounter. 
We keep track of this in our numerical calculations.

Photons IC scattered by the
jet when it is at a distance $d$ arrive at a far away observer
with a delay (wrt the arrival of first photons from the cocoon directly
without suffering any scattering) of
\begin{equation}
 t_{obs} \approx t_j + {d\over 2c\Gamma_j^2}.
\end{equation}
We are ignoring cosmological redshift factors in this and all other 
equations which can be easily included in the final expression for flux etc.

\begin{figure}[h!]
\centering
\includegraphics[scale=0.6]{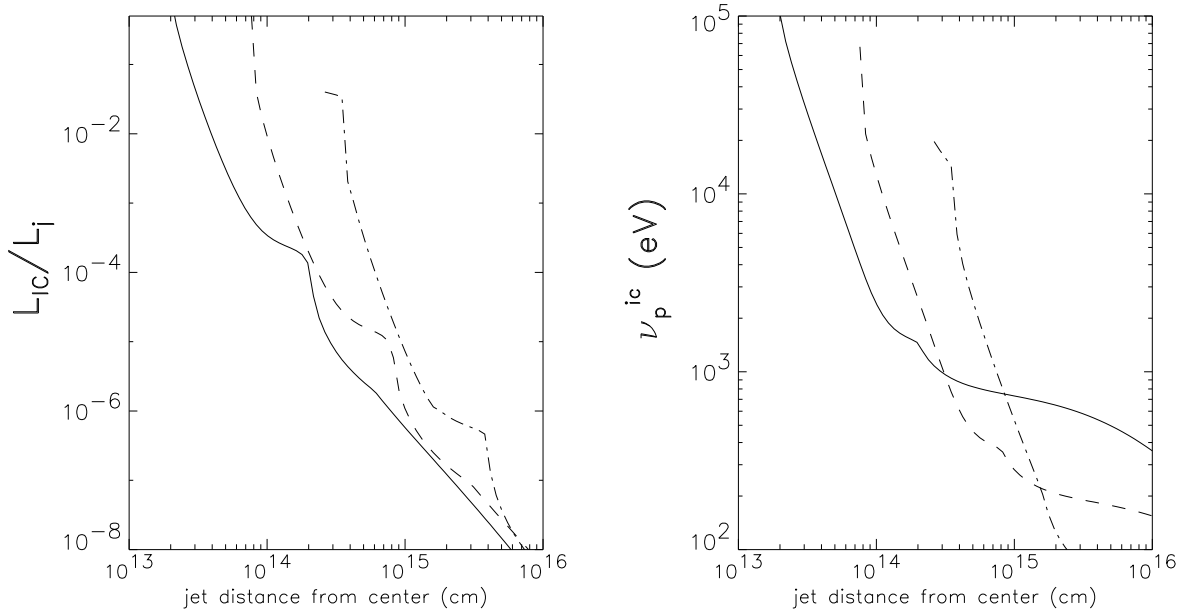}
\caption{The left panel shows IC luminosity divided by the luminosity
carried by the relativistic jet ($L_{ic}/L_j$) as a function of distance 
of the jet from the center of explosion. Electrons are taken to be cold 
in the jet comoving frame, i.e. $\gamma_e=1$; $L_{ic}\propto \gamma_e^2$
as long as the energy of photons from the cocoon as seen in electron rest 
frame is less than $m_e c^2$. The thermal energy in the 
cocoon is 10$^{52}$ erg (isotropic equivalent), its terminal Lorentz factor 
is 5, and the radius of GRB progenitor star is taken to be 10$^{11}$cm. 
The Lorentz factor of the relativistic
jet ($\Gamma_j$) is 100 for all calculations shown in this figure. And
its delay wrt to the time when cocoon punches through the stellar surface
($t_j$) is taken to be 10s (solid curve), 50s (dashed curve) and 250s
(dot-dash curve). We note that the x-axis can be converted to observer 
frame time using the equation $t_{obs} = [t_j + d/(2 c \Gamma_j^2)](1+z)$.
And to obtain the observed lightcurve we need to convolve the curve in
the left panel with $L_j(t)$, and include contributions to IC scatterings 
from parts of the jet moving at an angle larger than $\Gamma_j^{-1}$ wrt 
observer line of sight; the latter effect prevents lightcurve from falling
off faster than $[t_{obs}/(1+z) - t_j]^{-3}$ (Kumar and Panaitescu, 2000).
 The right hand panel shows the photon energy at the peak of the IC spectrum 
as a function of jet distance from the center for the same three values of 
$t_j$ as in the left panel.
   \label{ic-lum1} }
\end{figure}

\subsubsection{IC luminosity due to cocoon radiation scattering off of electrons in the jet}

The specific intensity of thermal radiation from the cocoon in its rest 
frame when it is at radius $r$ is $I'_{\nu'}(r)$. The specific intensity 
in the GRB host galaxy rest frame is related to $I'_{\nu'}$ as follows
\begin{equation}
  I_\nu(r) = I'_{\nu'}(r) \left[ { \nu\over \nu'}\right]^3 = {I'_{\nu'}(r) \over
    \Gamma_c^3 \left[ 1 - \beta_c\cos(\theta+\theta_1)\right]^3 },
\end{equation}
where
\begin{equation}
  \beta_c \equiv v_c/c, \quad\quad\quad \nu = {\nu'\over
     \Gamma_c \left[ 1 - \beta_c\cos(\theta+\theta_1)\right] } \equiv \nu'
     {\cal D}_1
   \label{doppler1}
\end{equation}
is the relativistic Doppler shift formula, and angles $\theta$ \& $\theta_1$
are defined in fig. \ref{fig2}. The bolometric thermal luminosity 
of the cocoon for a far away observer is 
\begin{equation}
  L_c^{iso}(r) = 4\pi d_A^2 \int d\Omega\int d\nu \, I_\nu = 2\pi^2 r^2 \nu'_p 
    I'_{\nu'_p} \Gamma_c^2, 
\end{equation}
where $\nu'_p$ is frequency at the peak of cocoon's thermal radiation spectrum.

The specific intensity in the jet comoving frame is given by
\begin{equation}
   I''_{\nu''} = I_\nu (\nu''/\nu)^3 = I_\nu \Gamma_j^3 (1 - 
   \beta_j\cos\theta_1)^3 = I'_{\nu'} \left[ {\Gamma_j(1 - \beta_j\cos\theta_1)
    \over \Gamma_c (1 - \beta_c\cos(\theta+\theta_1) } \right]^3,
    \label{Ij-trans}
\end{equation}

\begin{equation}
   \nu'' = \nu \Gamma_j(1 - \beta_j\cos\theta_1)\equiv \nu {\cal D}_2.
   \label{doppler2}
\end{equation}

Radiation from the cocoon is IC scattered by electrons in the jet.
If the Lorentz factor of electrons associated with their random motion in
the jet comoving frame is $\gamma_e$, then the bolometric IC luminosity
(isotropic equivalent) in observer frame when the jet is at a distance $d$
from the center of explosion is
\begin{equation}
   L_{ic}^{(iso)} = 4\pi d^2 (\gamma_e \Gamma_j)^2\int d\Omega_1''\int d\nu''\, 
    I''_{\nu''} \min\left[1, \tau_T(\theta_1) \right], 
  \label{Lic1}
\end{equation}
where 
\begin{equation}
    \tau_T(\theta_1) = {\sigma_T (L_j^{(iso)} \delta t_j)\over 4\pi d^2 m_p c^2 
    \Gamma_j \cos\theta'_1 \max\{1, c\delta t_j/(d/\Gamma_j^2)\} },
    \label{tau2}
\end{equation}
is the Thomson scattering optical depth of the causally connected part of 
the jet for photons moving at angle $\theta_1$ wrt to the jet axis;  
$\delta t_j$ is jet duration in GRB host galaxy rest frame, and 
\begin{equation}
    \cos\theta_1' = {\cos\theta_1 - \beta_j \over 1 - \beta_j\cos\theta_1}
\end{equation}
is cosine of the angle between jet axis and photon momentum in jet comoving 
frame.

The IC luminosity equation (\ref{Lic1}) can be rewritten in a more convenient
form by replacing jet comoving frame variable $\Omega''$ with 
host galaxy rest frame variable $\Omega$, and  $\nu''$ with $\nu'$:
\begin{equation}
 L_{ic}^{(iso)} = 4\pi d^2 (\gamma_e \Gamma_j)^2 \int d\Omega_1\int d\nu' \,
    I'_{\nu'} \min\left[1, \tau_T(\theta_1) \right] {\Gamma_j^2
   (1 - \beta_j\cos\theta_1)^2 \over \Gamma_c^4 [1 - 
     \beta_c\cos(\theta+\theta_1)]^4 },
    \label{L-ic1}
\end{equation}
where we made use of $d\Omega_1''/d\Omega_1 = {\cal D}_2^{-2}$, and
equations (\ref{doppler1}), (\ref{Ij-trans}) \& (\ref{doppler2}).

The peak of the IC spectrum in observer frame is at
\begin{equation}
   \nu_p^{(ic)} = {\Gamma_j \gamma_e^2 \int d\Omega_1''\,d\nu'' \,
    \nu'' I''_{\nu''} \min\left[1, \tau_T(\theta_1) \right]  \over
     \int d\Omega_1''\, d\nu'' \, I''_{\nu''} \min\left[1, \tau_T(\theta_1) 
   \right]} = {\Gamma_j \gamma_e^2 \int d\Omega_1\,d\nu' \,
    \nu' I'_{\nu'} \min\left[1, \tau_T(\theta_1)\right]{\cal D}_1^5{\cal D}_2^3
    \over
     \int d\Omega_1\, d\nu' \, I'_{\nu'} \min\left[1, \tau_T(\theta_1)\right]
   {\cal D}_1^4 {\cal D}_2^2 }.
\end{equation}

Numerical results for IC luminosity and peak photon energy are
shown in figure \ref{ic-lum1}, and order of magnitude estimates are
provided in the sub-section below. We point out that the IC luminosity
peaks roughly at the time when the jet emerges above cocoon photosphere,
and the peak value of $L_{ic}$ is of order the luminosity of the 
relativistic jet when $\Gamma_j \gae 100$, i.e. IC scatterings of 
cocoon radiation is very efficient in extracting energy from the jet at least 
for a brief period of time. The phase of high luminosity
lasts for a time of order the curvature time ($d/(2c\Gamma_j^2$) or jet 
duration (in GRB host galaxy rest frame) whichever is larger. The
IC luminosity decreases rapidly with jet distance from the center, 
as $\sim d^{-5}$, after the jet moves 
away from the cocoon photosphere. This decline is reduced to $d^{-2}$
for $d \gae R_{tr} (\Gamma_j/\Gamma_c) \max\{1, [t_j c/R_{tr}]^{1/2}\}$ when
 thermal photons are moving within an angle $\Gamma_j^{-1}$ of the jet axis 
and consequently IC scatterings do not boost the energy of photons; $R_{tr}$
is the radius where the cocoon becomes transparent to Thomson scatterings.
 The peak of the IC spectrum is roughly at 
$T_c (\Gamma_j/2\Gamma_c)^2\sim 10^5$ eV at the time when the jet emerges 
above the cocoon-photosphere (fig. \ref{ic-lum1}). 

\subsubsection{Order of magnitude estimate for IC luminosity}

A relativistic jet launched with a delay of time $t_j$, and moving 
outward at speed $v_j$, is at a distance $d$ from the center
of explosion at time $t_d = t_j + d/v_j\approx t_j + d/c + d/(2c \Gamma_j^2)$.
The jet is met with photons from the cocoon which were emitted in a certain 
region of $(r, \theta)$ plane at different times. 
The smallest cocoon radius ($R_{min}$) from which 
photons could arrive at the jet at time $t_d$ corresponds to 
$\theta=\pi/2$ and the maximum radius ($R_{max}$) is when $\theta=0$
(see fig. \ref{fig2}). Therefore, equations for the minimum and maximum
cocoon radii are obtained by equating jet and photon travel times to $d$,
 and are given by
\begin{equation}
 t_d = {R_{min}-R_*\over v_c} + {(R_{min}^2 + d^2)^{1/2}\over c} \approx 
   {R_{min} - R_*\over v_c} + {d\over c} \Longrightarrow
    R_{min} \approx R_* + v_c\left[ t_j + {d\over 2c \Gamma_j^2}
   \right],
\end{equation}
and
\begin{equation}
  t_d = {R_{max}-R_*\over v_c} + {d - R_{max}\over c} \Longrightarrow 
    R_{max} \approx
    2 (c t_j + R_*) \Gamma_c^2 + d(\Gamma_c/\Gamma_j)^2= 2\Gamma_c^2 (R_* +
   c t_{obs}).
\end{equation}
Photons emitted at $\theta> 1/\Gamma_c$ in the direction of the jet could 
get swept back up by the cocoon and scattered in a different direction before 
reaching the jet. This is certainly true when $R_{min}$ given by the above
equation is much smaller than $R_{tr}$. In that case a better lower
limit for cocoon radius from which photons can arrive at the jet is 
obtained by taking $\theta \approx 1/\Gamma_c$ and that gives
$R_{min} \approx t_{obs} v_c \Gamma_c^2$. 
The IC luminosity vanishes when $R_{min} > R_{tr}$ or $t_{obs} \gae 
R_{tr}/v_c$ since the cocoon luminosity drops off steeply for $r > R_{tr}$.

The equation for the curve in $(r, \theta)$ plane from which photons 
arrive at the jet at the same time is obtained by equating the time
for a photon from the cocoon to arrive at the jet and the time it takes for 
the jet to arrive at $r=d$ (including the launch delay of $t_j$), i.e.
\begin{equation}
    t_j = {r - R_* \over v_c} + {r_1\over c} - {d\over v_j} \approx {r - R_* 
    + r_1 - d\over c} + {r\over 2c\Gamma_c^2} - {d\over 2c\Gamma_j^2}.
   \label{time-delay0}
\end{equation}
This equation can be rewritten as
\begin{equation}
    t_j + R_*/c \approx {r\over 2c}\left( \theta^2 + \Gamma_c^{-2}\right) + 
   {r_1\over 2c} \left( \theta_1^2 - \Gamma_j^{-2}\right),
    \label{time-delay}
\end{equation}
where $r_1$, $\theta$ and $\theta_1$ are as defined in figure 
\ref{fig2}. Since, 
$r\theta\approx r_1\theta_1$, that means that the
time for the cocoon to travel to radius $r$ is a factor $\sim r_1/r$
larger than the time it takes for a photon to travel the distance $r_1$ to
the jet. We consider here the case where $d\ll 2ct_j \Gamma_j^2$, and for 
that the above time delay equation simplifies to
\begin{equation}
  t_j + R_*/c\approx {r\over 2c \Gamma_c^2}+ {r\theta^2 d \over 2c(d-r)} \approx
       {r\over 2c \Gamma_c^2}+ {(d-r)\theta_1^2 d \over 2c r}. 
   \label{time-delay1} 
\end{equation}
Considering that\footnote{This is because of forward beaming of photons 
from the cocoon to within an angle $\Gamma_c^{-1}$ of the cocoon's velocity
vector, and also because photons traveling at a larger angle wrt the radial 
direction will be swept back up and scattered in a different direction by 
the cocoon (if $r<R_{tr}$) before they can reach the jet.}
$\theta\lae \Gamma_c^{-1}$ we see from equation
(\ref{time-delay1}) that photons reaching the jet were emitted from
the cocoon when it was at a radius $r\approx\min\{2(c t_j+R_*)\Gamma_c^2,\, 
R_{tr}\}$ and at a latitude
\begin{equation}
  \theta \approx \left\{
\begin{array}{ll} \hskip -7pt
  \Gamma_c^{-1}  \quad\quad & t_j \ll R_{tr}/(2c\Gamma_c^2) \\  \\
  \hskip -7pt (2ct_j/R_{tr})^{1/2} \quad\quad &
                           t_j \gg R_{tr}/(2c\Gamma_c^2)
\end{array}
\right.
  \label{thetaa}
\end{equation}

The expression for IC luminosity (eq. \ref{L-ic1}) can be simplified using 
small angle and large Lorentz factor expansion, and is given by
\begin{equation}
 L_{ic}^{(iso)} \approx 32\pi^2 \gamma_e^2\Gamma_c^4 \int d\theta\, 
   \theta\, I'(r) r^2 \tau_T { \left[1 + (\theta_1\Gamma_j)^2\right]^2 \over 
     \left[ 1 + (\theta\Gamma_c)^2\right]^4 },
\end{equation}
where $I'(r)=L^{iso}_c(r)/(4\pi^2 r^2\Gamma_c^2)$ is the frequency-integrated 
specific intensity in cocoon comoving frame, $L_c^{iso}(r)$ is cocoon
luminosity (in host galaxy rest frame) when at radius $r$ which is 
given by equation (\ref{L_cocoon}), and $\theta$ \& $\theta_1$ (fig. 
\ref{fig2}) are related by $\theta_1 \approx r\theta/r_1 \sim r\theta/d$.
Most of the contribution to the integral in the above equation comes 
from $\theta\sim \Gamma_c^{-1}$ for $t_j \ll R_{tr}/(2c\Gamma_c^2)$, 
and $\theta_t\sim (2c t_j/R_{tr})^{1/2}$ when $t_j \gg R_{tr}/(2c\Gamma_c^2)$.
 The IC luminosity with this approximation is given by (as long as energies
of thermal photons from the cocoon are less than $m_e c^2$ in the comoving
frame of electrons in the jet)
\begin{equation}
 L_{ic}^{(iso)}(d) \approx \tau_T\gamma_e^2 \times \left\{
\begin{array}{ll} \hskip -7pt
  L_c^{iso}(r_t) \left[1 + (\theta_1\Gamma_j)^2\right]^2  \quad\quad  
                  & t_j \ll R_{tr}/(2c\Gamma_c^2) \\  \\
  \hskip -7pt 4 L_c^{iso}(r_t) (\theta_t\Gamma_c)^{-6} \left[1 + 
   (\theta_1 \Gamma_j)^2\right]^2 \quad\quad & 
                           t_j \gg R_{tr}/(2c\Gamma_c^2) 
\end{array}
\right.
  \label{L-ic1a}
\end{equation}
where
\begin{equation}
   r_t = \min\left\{R_{tr}, 2 c t_j \Gamma_c^2\right\}, \quad \theta_t\sim 
    \max\left\{\Gamma_c^{-1}, (2c t_j/R_{tr})^{1/2}\right\},
   \label{rt-thetat}
\end{equation}
and
\begin{equation}
  \theta_1 \approx \left\{
\begin{array}{ll} \hskip -7pt
  2c t_j\Gamma_c/d  \quad\quad & t_j \ll R_{tr}/(2c\Gamma_c^2) \\  \\
  \hskip -7pt (2ct_j R_{tr})^{1/2}/d \quad\quad &
                           t_j \gg R_{tr}/(2c\Gamma_c^2)
\end{array}
\right.
  \label{theta1a}
\end{equation}

According to eq. (\ref{tau2}) the optical depth of a relativistic jet to 
Thomson scatterings scales as $\tau_T\propto d^{-1}$ when the jet duration 
is longer than $d/(c\Gamma_j^2)$. In this case we see from equation 
(\ref{L-ic1a}) that $L_{ic}^{(iso)} \propto d^{-5}$ when $\theta_1\Gamma_j>1$, 
i.e. when the energy of IC scattered photons is larger than incident photon 
energy even for $\gamma_e=1$. At larger jet distances, when 
photons from the cocoon are traveling almost parallel to the jet axis in 
order to catch up with it, so that $\theta_1 \Gamma_j < 1$, the IC 
luminosity declines as $L_{ic}^{(iso)} \propto d^{-2}$. Numerical calculations 
confirm these rapid decline of IC luminosity with distance (fig. \ref{ic-lum1}).

The IC luminosity peaks at the time when the jet emerges above the 
cocoon surface where $d \sim r_t$ (eq. \ref{rt-thetat}), and $\theta_1 \sim 
\min(\theta_t, \Gamma_c^{-1})$. The maximum IC luminosity can be obtained 
from equations (\ref{L-ic1a}) and (\ref{theta1a}) and is given by
\begin{equation}
 L_{ic, max}^{(iso)}\approx \gamma_e^2 \times \left\{\begin{array}{ll} 
  \hskip -7pt L_c^{iso}(r_t) \tau_T(r_t) (\Gamma_j/\Gamma_c)^4 \quad\quad 
                  & t_j \ll R_{tr}/(2c\Gamma_c^2) \\  \\
  \hskip -7pt 4 L_c^{iso}(R_{tr}) \tau_T(R_{tr}) (\Gamma_j/\Gamma_c)^4 
      [R_{tr}/(2c t_j \Gamma_c^2)] \quad\quad &  t_j \gg R_{tr}/(2c\Gamma_c^2)
\end{array}
\right.
  \label{L-ic-max}
\end{equation}
The IC lightcurve peaks in the observer frame time at:
$t_{peak} \sim t_j + \min(r_t, R_{tr})/(2 c \Gamma_j^2)$, and the
temporal width of the peak is larger of the curvature time at $r_t$
($\approx r_t/2c\Gamma_j^2$) and the timescale for decline of relativistic 
jet luminosity.

The peak of the IC spectrum can be shown to be at a frequency 
\begin{equation}
  h\nu_p^{ic} \sim \gamma_e^2\Gamma_j \left[ {k_B T_c\over\Gamma_c}\right] 
  \left( {R_* \over r_t} \right)^{2/3} \left[ 
   { \Gamma_j(1 - \beta_j\cos\theta_1 \over \Gamma_c \{1 - \beta_c\cos(\theta_t
    + \theta_1)\}} \right] \sim {k_B T_c\gamma_e^2 [1 + (\theta_1\Gamma_j)^2] 
   \over 1 + [(\theta_t + \theta_1)\Gamma_c]^2 } \left[ {R_*\over r_t}
   \right]^{2/3}
  \label{ic-nu-p1}
\end{equation}
as long as scatterings are not in Klein-Nishina regime; here
$T_c$, $r_t$, $\theta_t$ and $\theta_1$ are given by equations 
(\ref{Tc}), (\ref{rt-thetat}) and (\ref{theta1a}). 
A more explicit expression for the IC peak frequency is obtained by 
substituting for these variables:
\begin{equation}
 h\nu_p^{ic}\sim \gamma_e^2 k_B T_c\left[{R_*\over R_{tr}}\right]^{2/3} \times 
   \left\{\begin{array}{ll}
  \hskip -7pt \left[1 + \left({ ct_j \Gamma_c\Gamma_j\over d}\right)^2
    \right]\left({R_{tr}\over R_* + 2c t_j \Gamma_c^2 }\right)^{2/3} \quad\quad
                  & t_j \ll R_{tr}/(2c\Gamma_c^2) \\  \\
  \hskip -7pt \left[ 1 + 2c t_j \Gamma_j^2 R_{tr}/d^2\right]  \left[ 1 + 2c t_j 
   \Gamma_c^2/R_{tr} \right]^{-1} \quad\quad &  t_j \gg R_{tr}/(2c\Gamma_c^2)
\end{array}
\right.
  \label{ic-nu-p2}
\end{equation}

\begin{figure}[ht!]
\centering
\includegraphics[scale=0.17]{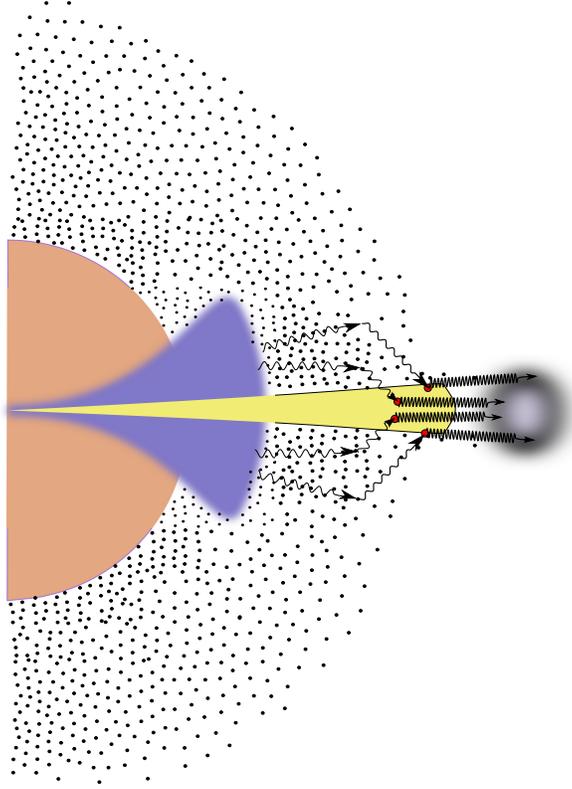}
\caption{Thermal photons from the cocoon are scattered by electrons in the
   wind that left the GRB progenitor star within the last few years of its 
   life. Some of these scattered photons collide with the relativistic 
   jet at a fairly large angle wrt the jet axis, and undergo strong
   inverse-Compton scattering by electrons in the jet. 
   \label{cocoon-wind-ic} }
\end{figure}

\subsection{IC scattering of wind-scattered cocoon radiation by relativistic jet}

Let us consider a spherical stream of photons produced by the cocoon moving
outward to larger radii. Some of these photons are scattered by electrons
in the circum-stellar medium --- wind from the GRB progenitor star
--- and arrive at the relativistic jet where they could suffer a second
scattering by electrons in the jet. We calculate the average photon energy 
and luminosity of these IC photons in observer frame.

The front of the photon stream from the cocoon is at radius
\begin{equation}
   r_{\gamma l} = R_* + t c,
\end{equation}
at time $t$ in the host galaxy rest frame, and its rear end is at
\begin{equation}
   r_{\gamma t} = \max\left\{ R_* + t v_c, \, R_{tr}+ (t - t_{tr})c \right\},
\end{equation}
where $R_{tr}$ is the radius where the cocoon becomes transparent to Thomson
scattering\footnote{The thermal luminosity of the cocoon drops rapidly
beyond the transparency radius ($R_{tr}$).} which is given by equation 
(\ref{r_ph_cocoon}), and
\begin{equation}
   t_{tr} = (R_{tr} - R_*)/v_c.
\end{equation}

The thermal flux from the cocoon at time $t$, and radius r between 
$r_{\gamma t}$ and $r_{\gamma l}$, is
\begin{equation}
   f_c(r,t) = {L_c^{iso}(r_1)\over 4\pi r^2},
\end{equation}
where $L_c^{iso}$ is given by equation (\ref{L_cocoon}), and 
\begin{equation}
    r_1 = R_* + \left[ r_{\gamma l}(t) - r\right] \beta_c/(1 - \beta_c).
\end{equation}

Let us consider electron density at radius $r$ associated with GRB progenitor 
wind to be
\begin{equation}
   n_e(r) = n_0 (R_*/r)^2.
\end{equation}

\begin{figure}[ht!]
\centering
\includegraphics[scale=0.2]{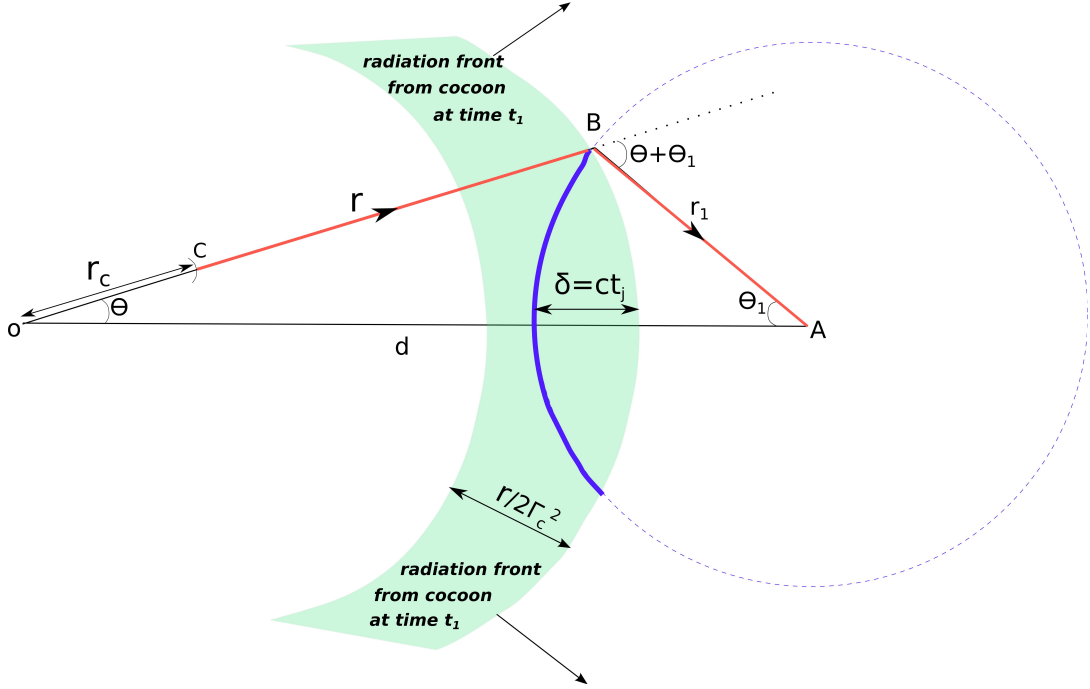}
\caption{A sketch showing a cocoon photon scattered by the
  CBM toward the jet. Radiation front from the cocoon at some time $t_1$ 
is shown in the shaded region. A photon was emitted at point $C$ from the
cocoon and travels to point $B$ where it is scattered by an electron 
in the CBM at time $t_1$ and arrives at the jet at point $A$ which is at a 
distance $r_1$ from point $B$. The highlighted segment of the circle of 
radius $r_1$ (that lies within the radiation front) are the set of all 
points from which photons scattered in the CBM at time $t_1$ can arrive 
at the jet at the same time. 
   \label{cbm-scat-sketch} }
\end{figure}

The IC luminosity calculation requires as input the specific intensity 
of wind-scattered thermal photons at the location of the jet.
Photons scattered in the wind at $(r, \theta, \phi=0, t_1)$ will arrive 
at the jet location $(d, 0, 0, t)$ provided that
\begin{equation}
  r = \left[ d^2 + (t-t_1)^2 c^2 - 2 c(t-t_1) d\, \cos\theta_1\right]^{1/2},
  \label{r_theta_t1}
\end{equation}
Where $\theta_1$ is the angle between jet axis and the photon momentum 
vector (fig. \ref{cbm-scat-sketch}), and is related to $\theta$ via the 
following equation
\begin{equation}
   r_1 \sin\theta_1 = r \sin\theta, \quad\quad r_1\equiv c(t - t_1).
\end{equation}

Equation (\ref{r_theta_t1}) can be solved to determine the time $t_1$ when
a photon at $(r, \theta)$ should be scattered so that it arrives at the
jet at time $t$. This in turn allows us to calculate the specific intensity
of wind-scattered photons at the location of the jet:
\begin{equation}
  \delta I_\nu^{(s)}(\theta_1) = \sigma_T f_c(r,t_1) n_e(r) \delta r_1/4\pi
\end{equation}
With specific intensity in hand we can calculate the IC luminosity using the
following equation
\begin{equation}
   L_{ic}^{(iso)} = 4\pi d^2 (\gamma_e \Gamma_j)^2\int d\Omega_1''\, d\nu''\, 
    \min\left[1, \tau_T(\theta_1) \right] I^{(s)''}_{\nu''},
\end{equation}
where $I^{(s)''}_{\nu''}$ is the specific luminosity of wind scattered photons
as measured in the rest frame of the jet. The equation for IC luminosity
can be rewritten in a more convenient form using Lorentz transformations of 
specific intensity, angle and frequency:
\begin{equation}
   L_{ic}^{(iso)}  = 4\pi d^2 (\gamma_e \Gamma_j)^2\int d\Omega_1\, d\nu\,  
    dr_1\, \min\left[1, \tau_T(\theta_1)\right] {dI^{(s)}_{\nu}\over dr_1}
   {\cal D}_2^2 ,
  \label{Lic2}
\end{equation}
where the Doppler factor ${\cal D}_2$ is defined in equation (\ref{doppler2}).

Numerical results for IC luminosity and peak photon energy are
shown in figure \ref{ic-lum2}, and order of magnitude estimates are
provided in the sub-section below. The IC luminosity
peaks roughly at the time when the jet emerges above the cocoon photosphere, and
then decreases as $\sim d^{-4}$ after the jet travels past 
the photosphere. 

\subsubsection{Order of magnitude estimate for IC luminosity}

Consider a photon that is emitted by the cocoon at radius $r_c(t_2)$ and
scattered by an electron in the CBM at radius $r$ toward the jet which is
at a distance $d$ from the center of explosion. The angle between the 
photon momentum and the jet axis is $\theta_1$ (fig. \ref{cbm-scat-sketch}).
 The requirement that photons arrive at the jet when it is at distance $d$ can 
be expressed as
\begin{equation}
   {r\theta^2\over 2} + {r_1\theta_1^2\over 2} + {r_c\over 2\Gamma_c^2} = c t_j
    + {d\over 2\Gamma_j^2},
  \label{delay2}
\end{equation}
where as before $t_j$ is the time when the jet is launched,
and other symbols are defined in figure \ref{cbm-scat-sketch}. 

At time $t$ the distance of the leading edge of the thermal radiation front 
from the center of explosion is given by
\begin{equation}
  r_{\gamma l}(t) = t c + R_* \approx t c,
\end{equation}
and the trailing edge is at
\begin{equation}
  r_{\gamma t}(t) = r_{\gamma l}(t) \left[ 1 - 1/2\Gamma_c^2\right].
\end{equation}
Photons near the leading edge left the cocoon when it was at radius
$R_*$ whereas photons near the trailing edge were produced close to
$r_{\gamma t}$. Therefore, for photons near the leading edge, the term 
$r_c/2\Gamma_c^2 \approx R_*/2\Gamma_c^2$ in equation (\ref{delay2}) can 
be neglected, and the equation simplifies to\footnote{We are also 
considering $d\ll 2c t_j\Gamma_j^2$ so that the second term on the right 
side of equation (\ref{delay2}) can be neglected.}
\begin{equation}
  t_j \approx {r \theta^2 d\over 2 c r_1} \approx {r_1 \theta_1^2 d\over 2 c r},
  \label{delay3}
\end{equation}
where we made use of a geometrical relation $r \theta = r_1\theta_1$ for the 
triangle OAB (fig. \ref{cbm-scat-sketch}). 

The intersection of a circle of radius $r_1$ centered at the jet head with 
the region containing cocoon radiation at time $t_1=t_j+d/v_j - r_1/c$ 
provides the locus of all points from which photons scattered at time
$t_1$ arrive at the jet at the same time (fig. \ref{cbm-scat-sketch}). 
The width of the overlap region of a circle of radius
$r$ (centered at the explosion site) and a second circle of radius $r_1$ 
(centered at the jet) is $\delta = r+r_1 - d \approx
r\theta^2/2 + r_1\theta_1^2/2 \approx c t_j$ (fig. \ref{cbm-scat-sketch}). 
Since the width of the radiation front is $r/2\Gamma_c^2$, all points on the
circle of radius $r_1$ that lie inside of the other circle are also
inside the radiation front at time $t_1$ as long as $r>2ct_j\Gamma_c^2$.
Therefore, the set of these points (shown in magenta color in fig. 
\ref{cbm-scat-sketch}) constitute the entire 1-D {\it hyper-surface} from 
which photons scattered in the CBM at $t_1$ arrive at the jet at the same 
time. From this little geometrical construction we see that the angular 
size of the beam of CBM scattered radiation that arrives at the jet at 
radius $d$ is equal to $\theta_1$ which is given by equation (\ref{delay3}).

The intensity of CBM scattered radiation is given by
\begin{equation}
   \delta I^{(s)} = {\sigma_T L_c^{iso}(r, t_1) n_e(r) \delta r_1\over 
   16\pi^2 r^2},
\end{equation}
and the IC luminosity due to scattering of these photons by the jet,
in the Thomson regime, is obtained from equation (\ref{Lic2}) using small 
angle expansion
\begin{equation}
   L_{ic}^{(iso)}  \approx {d^2 \gamma_e^2 \sigma_T\over 8} \int dr_1\,
   \int_0^{\theta_1} d\theta_a  {\tau_T L_c^{iso}(d-r_1, t_1) n_e(d-r_1)\over
    (d-r_1)^2} \left[ 1 + (\theta_a\Gamma_j)^2\right]^2 \theta_a,
   \label{Lic4}
\end{equation}
where $\theta_1$ --- the upper limit of $\theta_a$ integration --- is given 
by equation (\ref{delay3}).
It is easy to modify the above equation for $L_{ic}^{(iso)}$ to include 
Klein-Nishina cross-section when photon energy in electron rest frame is 
larger than $m_e c^2$.
For $d\gae 2ct_j\Gamma_c^2$ and $\theta_1\Gamma_j \gg 1$ equation (\ref{Lic4})
can be rewritten as:
\begin{equation}
   L_{ic}^{(iso)}  \approx {d^2 \gamma_e^2 \Gamma_j^4 \tau_T \sigma_T\over 48} 
    \int dr_1\,  {L_c^{iso}(d-r_1, t_1) n_e(d-r_1)\over (d-r_1)^2}
    \left[{2ct_j(d-r_1)\over r_1d}\right]^3.
   \label{Lic5}
\end{equation}
We note that the integrand for $L_{ic}^{(iso)}$ is a rapidly decreasing 
function of $r_1$ and therefore most of the contribution to the IC 
luminosity comes from the smallest possible value of $r_1$ which is 
$c t_j$ (this is to ensure that photons and jet arrive together at radius $d$ 
even though the jet was launched with a delay of $t_j$). 

We consider $d \gae 2c t_j \Gamma_c^2$ since at smaller distances the jet is 
below the cocoon's photosphere, and photons scattered in the CBM cannot
reach the jet (unless $d > R_{tr}$). Moreover, equation (\ref{Lic5}) is 
valid only for $d \lae 2c t_j \Gamma_j^2$, and so for 
$2c t_j \Gamma_c^2 \lae d \lae 2c t_j \Gamma_j^2$ the IC luminosity
is given by
\begin{equation}
   L_{ic}^{(iso)}  \approx (\tau_T d^2) {\gamma_e^2\Gamma_j^4 \over 
    12 } {\sigma_T L_c^{iso}(d) n_e(d) c t_j\over  d^2}.
    \label{Lic2c}
\end{equation}
This expression for IC luminosity has a simple physical interpretation. 
Since the integrand for $ L_{ic}^{(iso)}$ increases rapidly with decreasing
$r_1$ (eq. \ref{Lic5}) most photons arriving at the jet were scattered
by electrons in CBM within a distance $c t_j$ of the jet.
Hence the incident flux at the jet is $f_s\sim \sigma_T L_c^{iso} n_e ct_j/
d^2$ which is IC scattered by electrons in the jet to 
produce $L_{ic}^{(iso)}\sim f_s \tau_T \gamma_e^2 \Gamma_j^4 d^2$ as 
long as Klein-Nishina corrections are unimportant.

If the region around the polar axis of the star is evacuated by the
passage of a relativistic jet at an earlier time then there might be
a conical cavity of opening angle $\theta_m$ in the CBM containing few 
electrons to scatter cocoon photons toward the jet. In this case the 
lower limit for the radial integral in equation (\ref{Lic5}) is restricted to
\begin{equation}
  r_{1,min} = {\theta_m^2 d^2\over 2ct_j +  \theta_m^2 d}
\end{equation}
which follows from eq. \ref{delay3}), and therefore the IC luminosity
is given by
\begin{equation}
   L_{ic}^{(iso)}  \approx {\gamma_e^2\Gamma_j^4 \tau_T d^2 \over 12}
     {\sigma_T n_e(d) L_c^{iso}(R_{tr}) (c t_j)^3 \over r_{1,min}^2 d^2}.
    \label{Lic2d}
\end{equation}

\begin{figure}[ht!]
\centering
\includegraphics[scale=0.55]{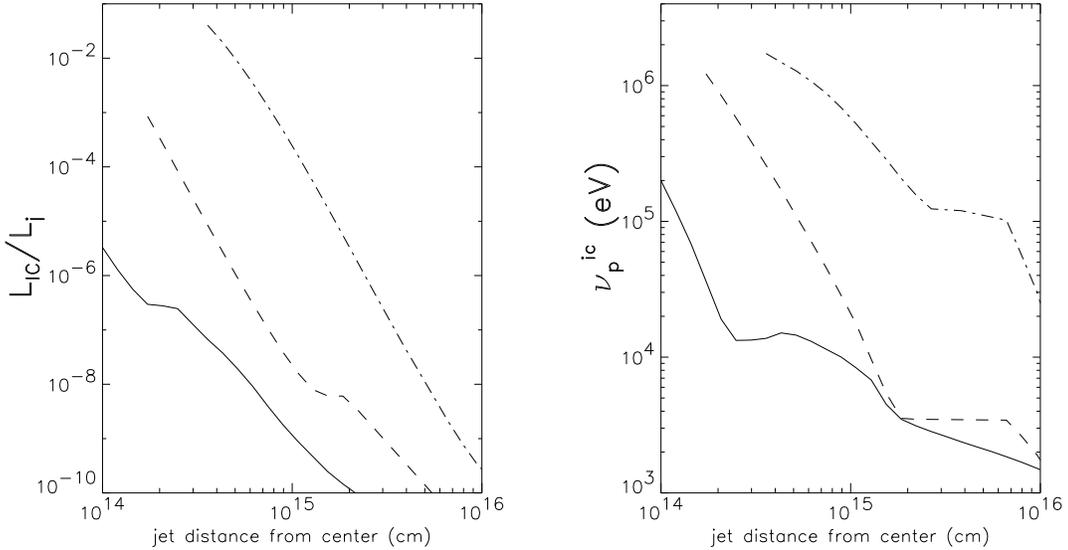}
\caption{The left panel shows IC luminosity divided by the luminosity
carried by the relativistic jet ($L_{ic}/L_j$) as a function of distance 
of the jet from the center of explosion. The difference between this and 
figure (\ref{ic-lum1}) is that here we consider thermal photons
from the cocoon to have been first scattered by electrons in the 
circum-burst medium (CBM) and a fraction of those run into the jet 
and undergo IC scattering whereas in the other case photons from the 
cocoon traveled to the jet directly.
The density of the CBM for these calculations is taken to be
$n_e(r) = 50\, r_{17}^{-2}$ cm$^{-3}$ which corresponds to mass loss rate of 
10$^{-5}$M$_\odot$ yr$^{-1}$, and wind speed of 10$^3$km/s, during the last 
10 years of the star's life; $r_{17}\equiv r/10^{17}$cm is the 
distance from the center of the star. A conical region of angular size 
0.1 rad along the jet axis is assumed to have been evacuated, i.e. $n_e=0$,
by the passage of an earlier relativistic jet, launched within a few 
seconds of the cocoon break-out, which passed through this region.
Electrons are taken to be cold in the jet comoving frame, i.e. $\gamma_e=1$;
$L_{ic}\propto \gamma_e^2$ as long as the energy of photons from the cocoon 
as seen in electron rest frame is less than $m_e c^2$. The thermal energy in 
the cocoon is assumed to be 10$^{52}$ erg (isotropic equivalent), its terminal 
Lorentz factor $\Gamma_c=5$, and the radius of GRB progenitor star is taken 
to be 10$^{11}$cm. The Lorentz factor of the relativistic
jet ($\Gamma_j$) is 100 for all calculations shown in this figure. And
its delay wrt to the time when cocoon punches through the stellar surface
($t_j$) is taken to be 10s (solid curve), 100s (dashed curve) and 10$^3$s
(dot-dash curve). 
 The right hand panel shows the photon energy at the peak of the IC spectrum
as a function of jet distance from the center for the same three values of
$t_j$ as in the left panel.
   \label{ic-lum2} }
\end{figure}

Equations (\ref{Lic2c}) and (\ref{Lic2d}) show that the IC luminosity
decreases with jet distance as $d^{-4}$ or faster which is consistent
with numerical calculations shown in fig. \ref{ic-lum2}; the luminosity peaks 
when the jet emerges just above the cocoon photosphere.

Another case we discuss is when the jet distance from the center of 
explosion is much larger than $2 c t_j\Gamma_j^2$. In this case the 
second term on the right side of equation (\ref{delay2}) dominates
and the various angles are given by
\begin{equation}
   \theta \approx \left({r_1\over r\Gamma_j^2}\right)^{1/2}, \quad\quad
      {\rm and} \quad\quad \theta_1 \approx \left({r\over r_1\Gamma_j^2}
   \right)^{1/2}.
   \label{delay5}
\end{equation}
Substituting this into equation (\ref{Lic4}) we find
\begin{equation}
   L_{ic}^{(iso)}  \approx {d^2 \gamma_e^2 \tau_T \sigma_T\over 16}
    \int dr_1\,  {L_c^{iso}(R_{tr}) n_e(d-r_1)\over (d-r_1)^2\Gamma_j^2}
    \left({d-r_1\over r_1}\right)\left[1 + {d-r_1\over r_1} + {(d-r_1)^2
    \over 3r_1^2} \right].
   \label{Lic7}
\end{equation}
For $\theta_1\Gamma_j \gg 1$ the integrand is proportional to $1/[(d-r_1) 
r_1^3]$, and therefore most of the contribution to IC luminosity comes from the
smallest possible value of $r_1$. A lower limit to $r_1$ is provided by
the consideration that the stellar wind within 
some angle, $\theta_m$, of the polar axis might have been swept-up
and evacuated by a relativistic jet that moved through the region before 
the current jet we are considering came along. Substituting $\theta=\theta_m$
into equation (\ref{delay5}) we obtain the following lower bound for $r_1$
\begin{equation}
  r_{1,min} = {\theta_m^2 \Gamma_j^2d\over 1 +  \theta_m^2 \Gamma_j^2}. 
\end{equation}
Finally, the IC luminosity when $\theta_1\Gamma_j = (d/r_{1,min}-1)^{1/2} \gg 
1$ is given by
\begin{equation}
   L_{ic}^{(iso)}  \approx {\gamma_e^2 \tau_T d^2 \over 48}
     {L_c^{iso}(R_{tr}) \sigma_T n_e(d) d \over 2\Gamma_j^2 r_{1,min}^2},
    \label{Lic2f}
\end{equation}
which declines with distance as $d^{-3}$.

\section{What can we learn from IC scattering of cocoon photons?}

Thermal radiation from cocoon can be IC scattered by GRB relativistic 
jets during the prompt $\gamma$-ray emission phase and also at later
times by jets associated with X-ray flares. The IC luminosity
depends on cocoon and jet properties, and also on the density of the
circum-burst medium\footnote{IC luminosity depends on circum-burst medium
density for the case where photons from the cocoon are first 
scattered by electrons in the CBM before bouncing off of the jet.}.
Therefore, IC photons could provide information regarding these 
different aspects of a GRB and its progenitor star.

A jet launched with a delay of $t_j$ emerges above the cocoon photosphere at 
a distance $r_t = \min\{2 c t_j \Gamma_c^2,\, R_{tr}\}$ from the center, and 
there it is bombarded with X-ray photons moving at an angle $\gae \Gamma_c^{-1}$
wrt jet axis. The IC scatterings of these photons by electrons in the 
jet produce high energy photons with a luminosity 
$L_{ic}^{(iso)}\sim L_c^{iso}\tau_T \gamma_e^2(\Gamma_j/\Gamma_c)^4$ 
(eq. \ref{L-ic-max}). The value for $L_{ic}^{(iso)}$ is of order the 
luminosity carried by GRB prompt relativistic jet even when electrons 
are cold in the jet frame (fig. \ref{ic-lum1}), and the energy of IC 
scattered thermal photons is a few hundred keV for $\Gamma_j=10^2$. 
Therefore, IC scatterings of cocoon photons by the GRB jet is an
important process that should be included in any discussion of 
$\gamma$-ray radiation mechanism and jet energetics\footnote{
The spectrum of IC scattered cocoon radiation is a Doppler broadened thermal
spectrum and not the usual Band function shaped GRB spectrum unless electron
distribution in jet comoving frame is a power-law function of its energy. 
Therefore, IC radiation is only a part of the observed GRB emission.}.

The IC luminosity declines rapidly with distance as the jet moves above the
cocoon photosphere ($L_{ic}^{(iso)}\propto d^{-5}$) since fewer and fewer 
cocoon-photons are able to catch up and collide with the jet at larger 
distances ($\theta_1\propto d^{-1}$; eq. \ref{theta1a}). 
Thus, a jet with a delay of $t_j$ provides information regarding 
cocoon radiation at a radius $\min\{2 c t_j \Gamma_c^2,\, R_{tr}\}$. 

If GRB jet energy is dissipated and electrons accelerated to high $\gamma_e$ 
somewhere within a radius of $\sim 10 r_t$ (so that $\gamma_e\Gamma_j \gae 
10^4$) then the IC radiation would peak at $\sim$GeV 
carrying a good fraction of the luminosity of the jet (fig. \ref{ic-lum1}). 
This process, therefore, should be useful for investigating jet dissipation 
and the poorly understood $\gamma$-ray radiation mechanism provided that 
$\gamma$-rays are produced within a radius of $\sim 10 r_t$.

Toma, Wu and Meszaros (2009) suggested that this process could 
explain a delay of a few seconds for GeV emission detected by Fermi/LAT
for a number of GRBs. However, that seems unlikely. Toma et al. considered 
radiation from the cocoon when it becomes transparent to Thomson scattering 
at a radius of $\sim10^{14}$cm and that resulted in a delay of a few seconds
for the IC emission.  However, the cocoon starts radiating soon after it 
breaks through the stellar surface and its luminosity declines monotonically 
after it stops accelerating and attains the terminal Lorentz factor of 
$\Gamma_c$ at $r\sim R_*\Gamma_c \sim 10^{12}$cm, and so the delay for 
the arrival of cocoon photons IC-scattered by the jet should be 
smaller than a few seconds. We showed in \S3 that IC luminosity peaks 
at the time when the relativistic jet emerges above the cocoon surface of 
optical depth 1 (and not the transparency radius) where it is bombarded 
with photons from the cocoon moving at an angle $\sim\Gamma_c^{-1}$ wrt 
the jet axis.  Moreover, the IC luminosity declines rapidly with distance
 ($d^{-5}$) as photons from the cocoon have to move increasingly parallel to 
the jet in order to catch up to it. In this regime, i.e. when 
$d \gg 2t_j c \Gamma_c^2$ --- which is what Toma et al. (2009) considered 
in their work --- one needs $\gamma_e\gg 10^2$ in order to obtain 
a significant IC luminosity. However, in this case the radiation produced 
within the jet --- which according to Toma et al. is observed as sub-MeV prompt
emission --- is much brighter than the radiation from the cocoon as 
measured in the jet comoving frame, and hence it is hard to see how the 
IC scattering of cocoon photons by the jet could be more important than 
scatterings of sub-MeV prompt $\gamma$-ray photons.

A detection of IC {\it quasi-thermal} component would provide information 
regarding cocoon luminosity and the ratio $\Gamma_j/\Gamma_c$.
These quantities are related to the structure of the outer envelope of GRB 
progenitor star. IC photons would also provide information regarding
electron thermal Lorentz factor in the region where prompt $\gamma$-ray
photons are produced provided that that takes place within $\sim 10 R_t$
as mentioned above.  IC radiation is polarized, and its measurement would 
shed light on GRB jet structure.

Even an upper limit on IC emission provides useful information. For instance,
the fact that $>$100 MeV emission from a typical GRB falls below Fermi/LAT
detection threshold suggests that sub-MeV $\gamma$-ray
prompt radiation is not produced between the distance of $\sim10^{12}$cm and
$10^{14}$cm from the center of explosion at least not involving a process that 
accelerates electrons to Lorentz factor larger than $\sim10^2/(\Gamma_j/100)$; 
$\gamma$-ray source radius of less than 10$^{15}$cm can be ruled out if
$\gamma_e\gae 10^3/(\Gamma_j/100)$ for those GRBs with high energy flux 
below the sensitivity of Fermi/LAT (fig. \ref{ic-lum1}). This result is 
useful for constraining the mechanism by which $\gamma$-rays are produced 
in GRBs. 

Late time jets ($t_j\gae10^2$s), such as those associated with X-ray flares,
 are useful for exploring CBM density. Cocoon photons scattered first by 
electrons in the CBM and subsequently IC scattered by late jets produce a bright
transient that peaks at a few x $\gamma_e^2$ MeV (fig. \ref{ic-lum2}). The flux 
is directly proportional to the density of the CBM which is 
related to the mass loss rate of GRB progenitor star during the last 
$\sim10$ years of its life.

GeV emission associated with late X-ray flares ($t_j \sim 500$ s)
is reported for a long duration GRB 100728A at redshift 1.57 
(Abdo et al. 2011). 
The X-ray and GeV emissions during the flare might be correlated, although 
the low photon statistics for Fermi/LAT precludes a definitive answer. 
The X-ray isotropic luminosity during the flare was $\sim 10^{49}$erg 
s$^{-1}$ which in the jet comoving frame is significantly smaller than the 
estimated cocoon's luminosity (fig. \ref{cocoon-prop}); X-ray flare data
suggest jet Lorentz factor to be larger than 30 (Abdo et al. 2011).
So from a theoretical point of view we expect cocoon photons to
scatter off of the late X-ray flare jet, both directly as well as
after bouncing off of CBM electrons, and produce high energy photons
with luminosity comparable to that in the X-ray band and that is consistent
with observations for this burst.

Another result of some interest is that the IC drag on
a jet composed of electron-positron pairs,
as opposed to electrons and protons, is so strong that the jet would lose
most of its energy soon after emerging above cocoon photosphere. Since that
is inconsistent with energy measured in GRB blastwave (from afterglow data)
we conclude that GRB jets cannot be dominated by $e^\pm$.

\section{Conclusion}

Relativistic jets of long duration GRBs push aside stellar material, and
evacuate a cavity, through the progenitor star on their way out to the surface.
This process creates a hot cocoon of plasma surrounding the jet with energy 
of order $10^{52}$ erg (isotropic equivalent). A fraction of this energy is 
radiated away on time scale of a few hundred seconds (in observer frame) 
when the cocoon punches through the stellar surface. The
interaction of this cocoon radiation with the relativistic jet has been 
investigated in this work and shown to be useful for exploring GRB jet 
and progenitor star properties. The basic idea is easy to explain. The
 radiative luminosity of the cocoon is of order 10$^{48}$ erg/s 
(isotropic equivalent), and photons from the cocoon collide with the jet 
at an angle of order $\Gamma_c^{-1}$ wrt jet axis; $\Gamma_c$ is the Lorentz 
factor of the cocoon. 
The cocoon luminosity as viewed in the jet comoving frame is a factor 
$\Gamma_j^2/(3\Gamma_c^4)$ larger. The rate at which radiation is 
produced by the jet --- which we see as prompt $\gamma$-ray or X-ray flare 
radiation --- is of order $10^{51}\Gamma_j^{-2}$ erg s$^{-1}$ (isotropic 
equivalent) in its comoving frame.  Therefore, the ratio of cocoon
thermal-radiation and jet radiation energy densities is 
$\sim 3\times10^{-4} (\Gamma_j/\Gamma_c)^4$ in the jet comoving frame. 
This ratio is larger than 1
for $\Gamma_j/\Gamma_c>10$, and in that case cocoon radiation is more 
important than the radiation produced within the relativistic jet for 
radiative cooling of electrons. 
Figures \ref{ic-lum1} and \ref{ic-lum2} show that the IC scattered cocoon 
radiation --- which forms a halo peaked near the edge of the jet and roughly 
as wide as the jet --- is of order the luminosity carried by the relativistic 
jet if electrons in the jet are heated to a thermal Lorentz factor larger than 
about 10$^2$ within a distance from the central engine of $\sim 10^{15}$cm.

The interaction of cocoon radiation with jet and predictions for high 
energy emission have been investigated in detail in this work. A lack of
detection of IC scattered cocoon thermal-radiation suggests either that the 
jet energy is not dissipated and imparted to electrons out to a radius of at 
least 10$^{15}$ cm --- which would rule out a certain class of models for GRB 
prompt emission --- or that the cocoon moves outward with a high Lorentz 
factor such that $\Gamma_j/\Gamma_c \lae 5$ (this possibility can be 
constrained by afterglow observations).

Photons from the cocoon scattered by electrons in the circum-burst medium
(CBM) can collide with the jet at a larger angle than photons traveling from 
the cocoon to the jet directly. These collisions result is very high energy 
photons ($\sim$GeV) of considerable luminosity even for a modest thermal 
Lorentz factor for electrons (fig. \ref{ic-lum2}). 
There is considerable uncertainty, however, in this estimate 
because the CBM could have been partially evacuated by an earlier passage
of a relativistic jet through this region. Detection of this signal, or 
an upper limit, would provide a handle on the stellar mass loss rate 
during the last few years of the life of the GRB progenitor star.

\medskip
\noindent{\bf Acknowledgment:} The work of GS is funded in part by PCCP.
"Paris Center for Cosmological Physics acknowledges the financial support 
of the UnivEarthS Labex program at Sorbonne Paris Cité (ANR-10-LABX-0023 
and ANR-11-IDEX-0005-02)".

\end{document}